\documentclass{article}
\usepackage{amsmath,amssymb,amsfonts}
\font\SYM=msbm10 
\newcommand{\Real}{{\SYM R}}
\newcommand{\Complex}{{\SYM C}}

\font\tenscr=rsfs10 scaled1100
\font\sevenscr=rsfs7 % scaled \magstep1 
\font\fivescr=rsfs5 % scaled \magstep1 
\skewchar\tenscr='177 
\skewchar\sevenscr='177
\skewchar\fivescr='177 
\newfam\scrfam 
\textfont\scrfam=\tenscr
\scriptfont\scrfam=\sevenscr 
\scriptscriptfont\scrfam=\fivescr

\def\scri{{\fam\scrfam I}}

\title{Conserved quantities in a black hole collision}
\author{S. Dain and J. A. Valiente-Kroon\\
  Max-Planck-Institut f\"ur Gravitationsphysik, Albert Einstein Institut\\
  Am M\"uhlenberg 1, 14476 Golm bei Potsdam, Germany}

\begin{document}

\maketitle

\begin{abstract}
  The Newman-Penrose constants of the spacetime corresponding to the
  development of the Brill-Lindquist initial data are calculated by
  making use of a particular representation of spatial infinity due to H.
  Friedrich.  The Brill-Lindquist initial data set represents the
  head-on collision of two non-rotating black holes. In this case one
  non-zero constant is obtained. Its value is given in terms of the
  product of the individual masses of the black holes and the square
  of a distance parameter separating the two black holes. This
  constant retains its value all along null infinity, and therefore it
  provides information about the late time evolution of the collision
  process. In particular, it is argued that the magnitude of the
  constants provides information about the amount of residual
  radiation contained in the spacetime after the collision of the
  black holes.
\end{abstract}

\bigskip

\emph{Introduction.} Friedrich and K\'{a}nn\'{a}r \cite{FriKan99a} have
calculated an expression for the so-called
Newman-Penrose\cite{NewPen68}  constants of
time symmetric spacetimes in terms of the initial data  on a
spacelike Cauchy hypersurface. The Newman-Penrose constants are
a set of 10 (5 complex) \emph{in general non-trivial} conserved
quantities defined as integrals of the form
\begin{equation}
G_m=\int_{S^2}\Psi_0^6\;\;{_2\overline{Y}_{2,m}} dS,
\end{equation}
over cuts of future null infinity ($\scri^+$), where
$\Psi_0=\Psi_0^5r^{-5}+\Psi_0^6r^{-6}+\cdots$ is 
the fastest decaying component of the Weyl tensor along the null
geodesics with affine parameter 
 $r$, ${_2Y_{2,m}}$ are spin-2 weighted spherical harmonics, and
$m=-2,\dots,2$. Generalisations of these constants can be constructed
to include the cases when null infinity is not smooth but
polyhomogeneous \cite{Val98,Val99a}.

The interpretation and physical meaning of these constants have been a
source of debate and controversy up to this day. They have been for
long time regarded as a mere mathematical curiosities, devoid of any
physical application. An explicit evaluation of these constants has
been carried out for stationary spacetimes \cite{NewPen68}. In this
case they were found to have the following structure:
\begin{equation}
G_m=(\mbox{dipole})^2-(\mbox{monopole})\times(\mbox{quadrupole}).
\end{equation}
This expression shows that the Newman-Penrose have a non-trivial
physical content even in the case of non-radiative spacetimes. Note
however, that the Newman-Penrose constants are all zero in the case of
the Schwarzschild spacetime. On the other hand, calculations of the
Newman-Penrose constants for non-stationary spacetimes are rather
scarse. The constants have been calculated for a class of type D
boost-rotation spacetimes \cite{LazVal00}.

The Friedrich-K\'{a}nna\'{a}r formula allows the evaluation of the
Newman-Penrose constants of a broad class of physically interesting
spacetimes, in particular certain space times  describing the head on collision of two
non-rotating black holes. This highly relativistic phenomenon is
usually studied by means of the numerical evolution of conformally
flat, time symmetric initial data. The initial 3-geometries are the
Brill-Lindquist data \cite{BriLin63} and the Misner data \cite{Mis63}.
The difference between the two initial data sets is essentially
topological. The Brill-Lindquist initial data possesses three
different asymptotically flat regions connected by two Einstein-Rosen
bridges, whereas the in the Misner data there are only two
asymptotically flat regions connected by a pair of bridges. The
physical implications of this difference of topology between the two
initial data sets is still not clear. Nevertheless, the expressions in
the Brill-Lindquist case are mathematically much simpler to handle
than those in the Misner case. Therefore, we 
concentrate our attention on the Brill-Lindquist data. However, a similar
calculations can be in principle carried out for the Misner data. The
evaluation of the Newman-Penrose constants of the Brill-Lindquist
initial data has a twofold objective. First, it seeks to gain some
insight on the physical interpretation of the
Newman-Penrose constants in the case of radiative spacetimes by
relating them to initial data quantities with a clear geometrical
meaning. And second, due to the fact that the Newman-Penrose constants
retain their value all along null infinity together with the
possibility of knowing their value directly from the initial data, one
may be able to extract information about the late time behaviour of the
complicate process of the collision of two black holes.

\emph{The Brill-Lindquist data.} As it has already mentioned, the
Brill-Lindquist initial data is time symmetric. Therefore, it is
solely in terms of the following negative-defined 3-metric:
\begin{equation}
ds^2=-\chi^4(dr^2+r^2d\sigma^2), \label{3metric}
\end{equation}
with $r=|x|$, $x\in \mathbb{R}^3$, and 
\begin{equation}
\chi=1+\frac{m_1}{2|x-x_1|}+\frac{m_2}{2|x-x_2|} \label{chi1},
\end{equation}
where $x_1$ and $x_2$ are two arbitrary points and $|\;\;\;|$ denotes the
euclidean distance. Without loss of generality we set the points $x_1$
and $x_2$ to lie along the $z$ axis. The origin of the coordinate system
is chosen be the middle point between the points $x_1$ and $x_2$.  
In the standard spherical coordinates, 
\begin{eqnarray}
&&|x-x_1|=\left(r^2+r_{12}\, r\cos\theta+(r_{12}/2)^2\right)^{1/2}, \\
&&|x-x_2|=\left(r^2-r_{12}\, r\cos\theta+(r_{12}/2)^2\right)^{1/2},
\end{eqnarray}
where $r_{12}=|x_1-x_2|$. Thus, using the generating function of the
Legendre polynomials one obtains directly the following expansions:
\begin{eqnarray}
&&\left(r^2+r_{12}r\cos\theta+(r_{12}/2)^2\right)^{-1/2}=\frac{1}{r}\sum^\infty_{n=0}(-1)^n\sqrt{\frac{4\pi}{2n+1}}Y_{n,0}\left(\frac{r_{12}}{2r}\right)^n,\\
&&\left(r^2-r_{12}r\cos\theta+(r_{12}/2)^2\right)^{-1/2}=\frac{1}{r}\sum^\infty_{n=0}\sqrt{\frac{4\pi}{2n+1}}Y_{n,0}\left(\frac{r_{12}}{2r}\right)^n,
\end{eqnarray}
which valid for $r>|x_1|$ and $r>|x_2|$ respectively. 
Hence, the scalar field $\chi$ defining the Brill-Lindquist 3-geometry is 
given by:
\begin{equation}
\chi=1+\frac{1}{r}\sum^\infty_{n=0}\sqrt{\frac{4\pi}{2n+1}}
\left(m_2+(-1)^n m_1\right)\frac{1}{2}\left(\frac{r_{12}}{2r}\right)^nY_{n,0}.
\label{chi}
\end{equation}
We are mainly concerned with the examination of the behaviour of these
initial data in the neighbourhood of the spatial infinity of one of
the 3 asymptotically flat regions. To this end, we make use of
Friedrich's \cite{Fri98a} representation of spatial infinity as a
cylinder $I=[-1,1]\times S^2$.  In order to do so, one has first to
compactify the 3-geometry. After introducing a new radial coordinate
$\rho=1/r$, one finds that the adequate conformal factor happens to be
$\theta^{-4}$ where
\begin{equation}
\theta = \chi/\rho.
\end{equation}
The resulting compactified 3-dimensional manifold is topologically
equivalent to the 3-dimensional sphere $S^3$. By construction we have
3 distinguished  points
in this compact manifold. These represent the infinities of the
initial data. All three spatial infinities are equivalent. When the
conformal metric is analytic, the conformal factor $\theta$ has near
$\rho=0$ the
following form\cite{Fri98a}:
\begin{equation}
\theta =\frac{U}{\rho}+ W,
\end{equation}
where $W$ and $U$ are analytic functions  of the appropriate
Cartesian coordinates. The function $U$ is
determined completely in terms of the local geometry near infinity.
Since the Brill-Lindquist data are conformally flat it follows that
$U=1$. The function $W$ contains information on the global geometry. In
particular, the total mass of the data at this end is given by
$2W(0)$. For the Brill-Lindquist data we have
$W=\rho^{-1}(\chi-1)$.

The cylinder of spatial infinity does not ``live'' in the unphysical
spacetime manifold but in a bundle with base manifold the conformally
compactified (unphysical) spacetime, and fibers given by
$CSL(2,\mbox{\Complex})=\mbox{\Real}^+\times SL(2, \mbox{\Complex})$.
In oder to make use of the results of \cite{FriKan99a} the angular
dependence in the function $W$ is rewritten in terms of the
functions $T_{m\;\;\;k}^{\;\;j}$, which are a complete orthogonal set
in $L^2(SU(2, \mbox{\Complex}))$. In particular, one has
\begin{equation}
Y_{n,0}=i^{2n}\sqrt{\frac{2n+1}{4\pi}}T_{2n \;\;\;\;n}^{\;\;\;n}.
\end{equation}
 Whence the lift of the function $W$ is given by:
\begin{equation}
W=\sum^\infty_{n=0}\frac{1}{2}\left( m_1+(-1)^n m_2\right)\left(\frac{r_{12}}{2}\right)^n\rho^n T_{2n \;\;\;n}^{\;\;\;n}.
\end{equation}
Therefore, we write
\begin{equation}
  \label{eq:W2}
  W= \sum_{n=0}^{\infty} \rho^n W_n,
\end{equation}
where
\begin{equation}
  \label{eq:Wn}
  W_n=  W_{n;2n,n}T_{2n \;\;\;\;n}^{\;\;\;n},
\end{equation}
and
\begin{equation}
  \label{eq:Wnn}
  W_{n;2n,n}=\frac{1}{2^{n+1}}(m_1+(-1)^n m_2)r^n_{12}.
\end{equation}
Note that the total ADM mass $m$ satisfies
$m=m_1+m_2=2W(0)=2W_{0;0,0}$. Friedrich \cite{Fri98a} has shown that
generic time symmetric data (like the Brill-Lindquist ones) give rise
to logarithmic divergences at the sets where $\scri^+$ ``touches'' the
cylinder $I$. In the same reference a regularity condition which
allows to avoid such divergences has been put forward. This condition
is given in terms of the Cotton tensor and its symmetrised
derivatives at spatial infinity. Now, the Brill-Lindquist initial data
is conformally flat, and thus it satisfies automatically the
regularity condition.  It should be remarked that it is still not
known whether the Brill-Lindquist 3-geometry develops a smooth null
infinity or not.  Nevertheless, the fact that the initial data
satisfy Friedrich's regularity condition is a good hint that a
smooth null infinity will be present. Here this  will be assumed to be 
the case.

\emph{The NP constants of the BL data.} If the spacetime is axially symmetric, then there is only one non-zero
Newman-Penrose constant and the Friedrich-K\'{a}nn\'{a}r formula
yields:
\begin{equation}
G_0=-\frac{1}{2}\sqrt{15\pi}\left\{127(\frac{1}{2}W_{0;0,0}W_{2;4,2}
-4W^2_{1;2,1})-\frac{1}{2\sqrt{6}}R_{2}\right\}, \label{janos}
\end{equation}
where $R_2=(\sqrt{6}/2)D_{(ab}D_{cd)_k}R$, and $R$ is the Ricci scalar
of the 3-metric on the initial hypersurface. If the initial data are 
locally conformally flat, then one can choose a conformal factor such
that $R=0$ in the neighbourhood of the reference asymptotic end. This 
is the case for the Brill-Lindquist data. 

Substituting the result of equation (\ref{eq:Wnn}) into equation
(\ref{janos}) one readily obtains the following remarkable result:
\begin{equation}
  \label{eq:fe}
  G_0=-\frac{127\sqrt{15\pi}}{4} r_{12}^2 m_1 m_2. 
\end{equation}
The quantity $G_0$ is  clearly of a quadrupolar nature. Note that if either
$r_{12}=0$ or any of $m_1$, $m_2$ are zero, one recovers the initial
data for the Schwarzschild spacetime, and consequently the constant $G_0$
vanishes. Even more interestingly, if the total mass $m=m_1+m_2$ is
kept fixed, then the constant maximizes its value in the case of the
most symmetric configuration, i.e. whenever $m_1=m_2=m/2$. Similar
results are expected to hold for the Misner initial data.

Assuming that the development of the Brill-Lindquist initial data
given by equation (\ref{chi}) gives rise to a smooth complete null
infinity then, the product given by equation (\ref{eq:fe}) will be
conserved along successive cuts of $\scri^+$. It is important to
observe that there are no other quantities  known  with such property.
Thus, this result may be useful to check the accuracy of numeric
simulations.

As a result of the head on collision of the two black holes, one
should expect the system to settle down to a Schwarzschild black hole.
The fact that the constant $G_0$ is zero only in the case where one
has a single black hole right from the beginning shows that this final
state of the evolution of the system is indeed an asymptotic state,
and that it cannot be reached in a finite amount of time. In other
words, there is always some amount of residual gravitation
radiational. Hence, the Newman-Penrose constants contain information
about the late time behaviour of the system. What is most remarkable
is the fact that this information is readily available from the
initial data!  In spacetimes which contain the point $i^+$ (future
timelike infinity, or equivalently past timelike infinity, $i^-$) the
Newman-Penrose constants have been interpreted as the value of the
Weyl tensor at $i^+$ \cite{NewPen68,FriSch87}. In a different context,
the Newman-Penrose constants have appeared as coefficients in the
leading term of late time expansions of the Bondi mass for some
boost-rotation symmetric spacetimes \cite{LazVal00}, therefore giving
a measure of ``how quickly the system is settling down to a
non-radiative state''.  Due to the non-trivial topology of the
Brill-Lindquist initial data, the conformal completion of the
spacetime resulting from the evolution of the Brill-Lindquist initial
data will not contain a regular  point $i^+$, however, one expects  a similar
interpretation to hold.  The formula (\ref{eq:fe}) suggests that an
initial configuration with the throats at a given separation should
contain more residual radiation than other configuration with the same
total mass but smaller separation. Analogously, a configuration with
similar masses should contain more residual radiation than another
with the same total mass but in which the ratio of the individual
masses is quite different from one. These ideas are in agreement with
recent numeric calculations for black-hole head on collisions
\cite{BakBruCamLouTak01,BakBruCamLou00}.

We thank Dr. H. Friedrich (MPI-AEI) and Dr. R. Lazkoz (Deustuko
Unibertsitatea, Spain) for careful readings of the manuscript and 
for several suggestions which lead to its significant improvement.

%\bibliographystyle{/afs/aei-potsdam.mpg.de/u/jav/tex/reporthack}
%\bibliography{/afs/aei-potsdam.mpg.de/u/jav/tex/thesisbib}

\end{document}